\documentclass[prb, reprint]{revtex4-1} 

\usepackage{amsmath}  
\usepackage{amsfonts}
\usepackage{color}

\newcommand{\caC}{{\mathcal C}}

\newcommand{\iu}{\mathrm{i}}

\newcommand{\str}{^{*}}

\newcommand{\Tr}{\mathrm{Tr}}

\newcommand{\braket}[2]{\left\langle #1 , #2\right\rangle}
\newcommand{\abs}[1]{\left\vert #1 \right\vert}

\newcommand{\ket}[1]{\left\vert #1\right\rangle}
\newcommand{\bra}[1]{\left\langle #1\right\vert}

\newcommand{\be}{\begin{equation}}
\newcommand{\ee}{\end{equation}}
\newcommand{\bea}{\begin{eqnarray}}
\newcommand{\eea}{\end{eqnarray}}
\newcommand{\beann}{\begin{eqnarray*}}
\newcommand{\eeann}{\end{eqnarray*}}

\begin{document}

\title[PVBS]{Product vacua with boundary states}

\author{Sven Bachmann}
\email{svenbac@math.ucdavis.edu}
\affiliation{Department of Mathematics\\
University of California, Davis\\
Davis, CA 95616, USA}

\author{Bruno Nachtergaele}
\email{bxn@math.ucdavis.edu}
\affiliation{Department of Mathematics\\
University of California, Davis\\
Davis, CA 95616, USA}

\date{\today}


\begin{abstract}
We introduce a family of quantum spin chains with nearest-neighbor interactions
that can serve to clarify and refine the classification of gapped quantum phases of such
systems. The gapped ground states of these models can be described as a product vacuum with a finite number of particles bound to the edges.
The numbers of particles, $n_L$ and $n_R$, that can bind to the left and right edges 
of the finite chains serve as indices of the particular phase a model belongs to.
All these ground states, which we call Product Vacua with Boundary States (PVBS) can
be described as Matrix Product States (MPS). We present a curve of gapped Hamiltonians 
connecting the AKLT model to its representative PVBS model, which has indices $n_L=n_R=1$. 
We also present examples with $n_L=n_R=J$, for any integer $J\geq 1$, that are related to a recently
introduced class of $SO(2J+1)$-invariant quantum spin chains. 
\end{abstract}


\maketitle


\section{Introduction}
\label{sec:introduction}

Gapped quantum phases and the transitions between them are a subject of
great current interest both for their fundamental importance, as illustrated by the experimental 
observation of an $E_8$-symmetry at the critical point of the quantum Ising model~\cite{Coldea}, and because of their potential application in quantum information and 
computation as proposed in~\cite{Kitaev}. In the latter
case the focus is on the 
topological phases \cite{topo1, *Wen1990}, which cannot 
be classified using a local order parameter,
but depend on global features of 
the system such as the underlying topology of  the model, as for example in the 
quantum Hall effect \cite{laughlin}. These topological orders can 
be characterized by an equivalent boundary theory \cite{Halperin, *Juerg, *WenEdge}.

Several recent works have aimed at classifying the ground state phases of quantum spin chains
\cite{VerstraeteWolf, *yoshida, WenSymm}. While true 
topological order may not occur in one dimension,
direct analogues of some essential phenomena are present, including the role of edge states,
which is the main topic of this article.

The consensus is that the ground states of two models $H_0$ and $H_1$
with short-range interactions are in the same phase if there exists a smooth path of Hamiltonians $H(s)$, $0\leq s\leq 1$, $H_0=H(0), H_1=H(1)$, such that the gap 
above the ground state does not close along this path \cite{WenHastings, *ChenGuWen, BMNS}. 
Using the notion of automorphic equivalence, in \cite{BMNS} we 
showed how this definition can be made precise so that it can also be applied for infinite systems. 
In general, proving that the gap does not close is a hard problem but a criterion exists
for frustration free models whose ground states are Matrix Product States (MPS) \cite{VBSHam}. 
Most progress has been made by focusing on such frustration free models, with or without 
prescribed symmetries \cite{WenSymm, schuch}. 

In this article, we introduce a new family of quantum spin chains with nearest-neighbor interactions
with MPS ground states of a special type which we call Product Vacua with Boundary States 
(PVBS). The bulk phase is a product state (or several product states), but for finite chains 
there are $2^n$ ground states that are obtained from the product vacuum by adding up to $n$ 
distinguishable particles. Of these $n$ particles $n_L$  bind to the left edge of the interval
and $n_R$ to the right edge. The equivalence classes of gapped ground states are in one-to-one
correspondence with the values of the non-negative integers $n_L$ and $n_R$. The Hamiltonian 
preserves the particle number for each type of particle separately. As an example, we identify 
the classes to which the $SO(2J+1)$-invariant models of~\cite{SOn} belong, namely $n_L=n_R=J$. The AKLT model corresponds to $J=1$. The complete class of PVBS models
is more general and will be described elsewhere.


\section{Product vacua with boundary states}
\label{sec:pvbs}

We consider a quantum spin chain with spin dimension $d\geq n+1$. There are $n+1$ states 
at each site that we interpret as $n$ distinguishable particles and a vacuum. In addition there may
be excited states of positive energy that will play no role in our discussion. For simplicity, we stick to the case $d=n+1$. We start from a MPS representation of the ground states of the model. The Hamiltonian is constructed as the parent Hamiltonian for the set of MPS ground states and by general arguments we can then conclude that it has a gap and no 
other ground states \cite{FCS}. 
 
Let $0$ label the empty state and let 
$1,\ldots,n$ denote
the $n$ particle types. The ground states are generated by $n+1$ square matrices
$v_0,v_1,\ldots,v_n$, satisfying the following commutation relations:
\begin{eqnarray}
v_i v_j&=&e^{i\theta_{ij}} \lambda_i \lambda_j^{-1} v_j v_i, \quad i\neq j\label{cr1}\\
v_i^2 &=& 0, \quad i\neq 0\label{cr2}
\end{eqnarray}
where $\theta_{ij}\in\mathbb{R}$, $\theta_{ij}=-\theta_{ji}$, and $0\neq \lambda_i\in \mathbb{R}$, 
for $0\leq i,j\leq n$. By redefining the phases $\theta_{ij}$, we
can assume $\lambda_i >0$ without loss of generality.
We will also assume that $\lambda_0 =1$, which amounts to a choice of normalization for $v_0$.
Although the exact form of the matrices is not essential for our discussion, it is
important that such matrices exists. One representation of the commutation relations is most conveniently described in terms of a chain of $n$ spin-$1/2$ particles, yielding matrices of dimension $2^n$.
Define 
\begin{eqnarray*}
\sigma^+ =& \begin{pmatrix} 0 & 1 \\ 0 & 0 \end{pmatrix} \qquad
\sigma^- =& \begin{pmatrix} 0 & 0 \\ 1 & 0 \end{pmatrix} \\
w_i =& \begin{pmatrix} 1 & 0 \\ 0 & \lambda_i \end{pmatrix}  \qquad
P_{ij} =& \begin{pmatrix} e^{i\theta_{ij}/2} & 0 \\ 0 & 1 \end{pmatrix}
\end{eqnarray*}
for $i,j=0,\ldots n$. The matrices are given by
\begin{eqnarray*}
v_0 &=& \bigotimes_{i=1}^n P_{0i}^2w_i\,, \\
v_i &=& \bigotimes_{j=1}^{i-1} P_{ij}w_j \otimes \sigma^+ \otimes \bigotimes_{k=i+1}^n P_{ik}w_k\,,\quad i=1,\ldots n\,.
\end{eqnarray*}
Eq.~(\ref{cr1}) follows from $\sigma^+w_i = \lambda_i w_i \sigma^+$ and $P_{ij}\sigma^+ = e^{i\theta_{ij}/2}\sigma^+P_{ij}$ as well as $[P_{jk},w_i] = 0$ for all $i,j,k$. Finally,~(\ref{cr2}) is a direct consequence of $(\sigma^+)^2 = 0$.

The MPS generated by this set of matrices for a chain of $L$ spins are given by
\begin{equation}
\psi(B) = \sum_{i_1,\ldots,i_L=0}^n \Tr (B v_{i_L}\cdots v_{i_1})\ket{i_1,\ldots,i_L}\,,
\label{mps}
\end{equation}
where $B$ is an arbitrary $2^n\times 2^n$ matrix. Consider the case $L=2$.
From the commutation relations (\ref{cr1}) and (\ref{cr2}) it follows that all MPS vectors of
the form (\ref{mps}) will be orthogonal to the vectors 
$\phi_{ij}\in \mathbb{C}^d\otimes\mathbb{C}^d$ given by
\begin{eqnarray}
\phi_{i}&=&\ket{0,i}-e^{-i\theta_{i0}}\lambda_i \ket{i,0} \label{phii}\\
\phi_{ij}&=&\ket{i,j}-e^{-i\theta_{ji}}\lambda_i^{-1} \lambda_j\ket{j,i} \label{phij}\\
\phi_{ii}&=&\ket{i,i} \label{phiii}
\end{eqnarray}
for $1\leq i,j \leq n$ and $i\neq j$. If $n>2$, not all combinations of 
particles can be realized in a ground state of a chain of length $2$.
Nevertheless, it is sufficient to consider a Hamiltonian with a nearest-neighbor interaction
defined by
$$
h=\sum_{i=1}^n |\hat{\phi}_i\rangle\langle\hat{\phi}_i| + \sum_{1\leq i \leq j\leq n}^n |\hat{\phi}_{ij}\rangle\langle\hat{\phi}_{ij}|,
$$
where $\hat{\cdot}$ denotes normalization.
The Hamiltonian for a finite chain of spins is then given by
\begin{equation}
H_{[a,b]}=\sum_{x=a}^{b-1} h_{x,x+1},
\label{ham}\end{equation}
where and $h_{x,x+1}$ is a copy of $h$ acting on the pair of spins at the sites
$x$ and $x+1$. As a sum of orthogonal projections, $H_{[a,b]}$ is non-negative and it is straightforward to verify that the MPS defined in (\ref{mps}) are eigenvectors with
zero energy, hence ground states of the model. It is not hard to show that these are {\em all} the ground states of $H_{[a,b]}$ \cite{FNW_ff}.

If $b-a+1\geq n$, for each subset $\{i_1,\ldots,i_m\}$ of $\{1,\ldots,n\}$ there is a ground state 
$\psi_{[a,b]}^{i_1,\ldots,i_m}=\psi(B^{i_1,\ldots,i_m})$ having exactly one particle of each type $i_1,\ldots,i_m$. The matrices $B^{i_1,\ldots,i_m}$ generating $\psi^{i_1,\ldots,i_m}$ through~(\ref{mps}) (up to normalization) can be chosen as
\begin{multline*}
B^{i_1,\ldots,i_m} = p^{\otimes (i_1-1)}\otimes \sigma^-\otimes p^{\otimes (i_2-i_1-1)}\otimes \sigma^- \otimes \\
\cdots \otimes\sigma^-\otimes p^{\otimes (n-i_m)}
\end{multline*}
where $p = \sigma^+\sigma^-$. An interesting example is the ground state containing only particle $i$, 
\begin{equation}
\psi_{[a,b]}^i = \sum_{x=a}^b \left(e^{i\theta_{i0}}\lambda_i\right)^x \ket{0,\ldots,i,\ldots,0}
\label{one_particle}\end{equation}
where $i$ is at site $x$ in each term of the sum.  We now add the assumption that $\lambda_i\neq 1$, for $1\leq i\leq n$.
The $n_L$ particles having $\lambda_i<1$ are bound to the left edge, whereas the 
$n_R = n-n_L$ particles with $\lambda_i>1$ are localized near the right edge, as can be seen in 
(\ref{one_particle}). The state with $m=0$ is the product state
\begin{equation*}
\Omega = \ket{0,\ldots,0}\,.
\end{equation*}
All other ground states differ from $\Omega$ only near the edges. Specifically,
\begin{equation*}
\lim_{\substack{a\to-\infty \\ b\to+\infty}}\braket{\hat{\psi}_{[a,b]}^{i_1,\ldots,i_m}}{ A 
\hat{\psi}_{[a,b]}^{i_1,\ldots,i_m}} = \braket{\Omega}{ A \Omega}
\end{equation*}
for any local observable $A$. If only one of the edges is taken to infinity, the limiting ground states 
for the half-infinite chain depend on the particles at the other edge. Concretely, on the chain that extends to infinity on the right but with a left boundary, there remain $2^{n_L}$ ground states corresponding to the possible combinations of the $n_L$ particles that bind to the left edge. Similarly, there are $2^{n_R}$ ground states on the left infinite chain with a right boundary.

Using the method of \cite{VBSHam}, we can prove that the energy of the first excited state is 
bounded below by a positive constant, independently of the length of the chain. As at most one 
particle of each type can bind to the edge, any second particle of that type must be in a scattering 
state. The dispersion relation of these scattering states can be explicitly calculated by considering the restrictions of the Hamiltonian to any of the $n$ invariant spaces containing exactly one particle. Properly centered and rescaled, these operators reduce to a free hopping Hamiltonian with $\lambda$-dependent boundary conditions. A plane wave Ansatz yields the dispersion relation
$$
\epsilon_i(k) = 1 -\frac{2\lambda_i}{1+\lambda_i^2}\cos(k+\theta_{i0})\,.
$$
In particular the gap closes whenever $\lambda_i\to1$ for some $i$. Moreover, we conjecture that the \emph{exact} gap in the thermodynamic limit is given
by 
$$
\gamma=\min\left\{ \frac{(1-\lambda_i)^2}{1+ \lambda_i^2}: 1\leq i\leq n\right\}\,.
$$


\section{Automorphic equivalence and gapped ground state phases}
\label{sec:auto}

Despite their simplicity, the PVBS models introduced in the previous section
display the general characteristics of gapped one-dimensional systems with a unique bulk ground state and we can use them to illustrate the role of edge states in the classification of gapped phases.

In~\cite{WenUnitary} it is concluded that all gapped, translation 
invariant, one-dimensional quantum spin systems without symmetry breaking belong to the same 
phase, and that they are equivalent to a product state. We believe that taking into 
account edge states, and in particular the behavior of the system on semi-infinite chains,
is necessary as it allows for a finer classification closer in spirit to what one would find 
 for models on two- and higher dimensional manifolds with nontrivial topology.
 

\subsection{The PVBS classes}
\label{sub:PVBS}

It is easily seen that two PVBS models of the type we introduced here
belong to the same equivalence class if and only if they have the same values
for the  non-negative integers $n_L$ and $n_R$. The reasoning is as follows.
Since equivalent phases are related by an automorphism, a unique bulk 
ground state can only be mapped to another unique bulk state. Similarly, the
ground state space dimensions of the half-infinite chains, $2^{n_L}$ and 
$2^{n_R}$, are also preserved by an automorphism. 
Hence, if two PVBS models belong to the 
same phase, they must have equal $n_L$ and $n_R$. Conversely, if two PVBS models 
have the same values of $n_L$ and $n_R$ but each with their own sets of parameters
$\{\lambda_i(\alpha)\mid 1\leq i\leq n_L+n_R\}$
and $\{\theta_{ij}(\alpha)\mid 1\leq i,j\leq n_L+n_R\}$, for $\alpha=0,1$,
an interpolating path along which the gap does not close can be constructed as follows. 
First,  one may apply a strictly local unitary to perform a change of basis in spin space such
that both are PVBS expressed in the same spin basis and such that $\lambda_i(\alpha) <1$
for $1\leq i\leq n_L$ and $\lambda_i(\alpha) >1$ for $n_L+1\leq i\leq n_L+ n_R$, for both
$\alpha=0$ and $\alpha=1$. Let $u$ be the unitary for this change of basis. Then, take
a smooth curve of unitaries $u(s)$, $0\leq s\leq 1$, with $u(0) = \openone$ and $u(1) = u$, and let $U_{[a,b]}(s)$ be the $(b-a+1)$-fold tensor product of  $u(s)$. 
Now we conjugate the initial $H_{[a,b]}$ with these unitaries to define a smooth path of Hamiltonians with a constant gap.
Simultaneously, we can deform the parameters of the two models by linear interpolation:
\begin{eqnarray*}
\lambda_i(s) &=& (1-s)\lambda_i(0) + s\lambda_i(1)\label{lambdas},\\
\theta_{ij}(s) &=& (1-s)\theta_{ij}(0) + s\theta_{ij}(1)\label{thetas}.
\end{eqnarray*}
This yields a smooth family of vectors $\phi_{ij}(s)$ as in~(\ref{phii}-\ref{phiii}) and thereby a smooth 
family of nearest-neighbor interactions $h(s)$ and of Hamiltonians. The gap remains open
because $\lambda_i(s)\neq 1$ for all $1\leq i \leq n$ and $s\in [0,1]$.
By the general result of \cite{BMNS} this implies the quasi-local automorphic equivalence 
of the two models. Note that it is essential that in each pair between which we interpolate,  the
$\lambda_i$'s are either both $<1$ or both $>1$, which is why we had to assume that
$n_L$ and $n_R$ are the same for both models. If one uses the same type of interpolation
to connect models with different values of $n_L$ and $n_R$, the gap necessarily closes
along the path and there is a quantum phase transition.
This is not to say that one could not construct paths along which the gap closes also
for the case of constant $n_L$ and $n_R$ and of course this would then not imply a
transition between different gapped phases.

The uncountable family of PVBS models which depend on the real parameters $\{\lambda_i\}$ and $\{\theta_{ij}\}$ is completely classified by the pair of integers $(n_L,n_R)$. Given their simplicity, it is natural to choose them as representatives of the much larger phase they belong to.


\subsection{The AKLT model}
\label{sub:AKLT}

As a first example, we show that the AKLT model \cite{AKLT} belongs to the same
equivalence class as the PVBS models with $n_L=n_R=1$. The AKLT model
is an antiferromagnetic spin-$1$ chain with a unique, gapped ground state in the
thermodynamic limit, and four zero-energy states on a finite chain, which are usually 
described in terms of a spin-$1/2$ particle attached to the two ends of the chain.
We found a smooth path of gapped models connecting the AKLT model with a PBVS model
with one particle for each boundary. Let us denote the two
particle states by $-$ and $+$. For $s\in[0,s_0]$ where $\sin(s_0)=\sqrt{2/3}$, the following
four vectors span the ground state space of two neighboring spins of the the interpolating
models as a function of $s$:
\begin{eqnarray*}
\psi^{00}(s) &=&  \mu(s) \sin (s) \left[\lambda(s)^2 \vert -,+\rangle + \vert +,-\rangle \right]\\
&&- \cos^2(s)(1+\lambda(s)^4) \vert 0,0\rangle \\
\psi^{0-}(s) &=& -\lambda(s)\vert 0,-\rangle + \vert -,0\rangle \\
\psi^{0+}(s) &=& - \lambda(s)\vert +,0\rangle + \vert 0,+\rangle\\
\psi^{-+}(s) &=& \vert -,+\rangle - \lambda(s)^2\vert +,-\rangle \,,
\end{eqnarray*}
where $\lambda(s)$ is a smooth function such that $\lambda(s_0) = 1$, $0<\lambda(s)<1$, for all 
$s<s_0$, and $\mu(s) = (1-\lambda(s)^2 \cos^2(s))^{1/2}$. The corresponding
nearest-neighbor interaction,
\begin{equation*}
h(s) = 1 - \sum_{(ij)\in\{(00), (0-), (0+), (-+)\}}\ket{\widehat{\psi^{ij}}(s)}\bra{\widehat{\psi^{ij}}(s)}
\end{equation*}
is the 
projection onto the orthogonal complement of this $4$-dimensional space. 
We then define $H_{[a,b]}(s) = 
\sum_{x=a}^{b-1}h_{x,x+1}(s)$. It readily follows that $H_{[a,b]}(s_0)$ is the AKLT Hamiltonian and 
that $H_{[a,b]}(0)$ is the PVBS model with $n_L=n_R=1$, the coefficients $\lambda_- = 
\lambda(0)$ and $\lambda_+ = \lambda(0)^{-1}$, and all the phases $\theta_{ij} = \pi$. The path 
of interactions is smooth as the four ground state vectors are smooth, remain orthogonal 
to each other and of finite norm for all $s$, and the spectral gap does not close~\cite{Prep}. 
Hence, the AKLT model belongs to the same gapped quantum phase as the PVBS model with $n_L=n_R=1$. In particular, the 
sets of ground states of these models are automorphically equivalent for the finite, 
half-infinite and infinite chains, where they are isomorphic to a pair of qubits, a single qubit, 
and a unique pure product state, respectively.

The ground states of $H_{[a,b]}(s)$ for 
$s>0$ have a minimal matrix product representation with $2\times 2$ matrices $v_i(s)$ defined
as follows:
\begin{eqnarray}
v_0 &=& -\cos(s)\begin{pmatrix} 1 & 0 \\ 0 & -\lambda(s) \end{pmatrix}, \label{v0s} \\
v_- &=& \begin{pmatrix} 0 & -\mu(s) \\ 0 & 0 \end{pmatrix},  \quad v_+ = \begin{pmatrix} 0 & 0 \\ 
\sin(s) & 0 \end{pmatrix}. \label{vpms}
\end{eqnarray}
By contrast, the MPS representation of the ground states of the PVBS model, at $s=0$, uses 
$4\times4$ matrices. In fact, there is no faithful $2$-dimensional representation of the commutation 
relations~(\ref{cr1}, \ref{cr2}) with $n=2$. Up to unitary change of basis, the only nilpotent matrix is 
$\sigma^+$, so we can set $v_+=c_+\sigma^+$ and $v_- = c_-U\str \sigma^+ U$, with 
$c_\pm\neq0$ and $U\neq1$. But their commutation relation implies $\mathrm{Tr}(v_-v_+)=0$ 
so that $U=1$ if $c_\pm\neq 0$. The parent Hamiltonian constructed from the 
limiting matrices  $v_i(0)$ corresponds to a PVBS model with $n_L=1$ and $n_R=0$.
With a minor modification one can interchange the roles of $n_L$ and $n_R$, but
in either case the limiting matrices lead to a parent Hamiltonian that is {\em not} 
in the same phase as the AKLT model; It turns out to be equivalent only in the bulk
and on the half-infinite chain tending to $\infty$ on the right but not on the chain that is infinite
on the left. We believe that the finer classification we propose in this paper yields interesting 
additional information compared to previous approaches such as \cite{schuch}. 
We conclude that it is essential to construct a smooth, gapped family 
of Hamiltonians, and that it is not sufficient to find a smooth family of MPS matrices only. A similar 
conclusion was reached in~\cite{schuchetal}, where `uncle Hamiltonians' were constructed that share 
the ground state spaces of gapped `parent Hamiltonians' but are gapless in the thermodynamic limit.


\subsection{The $SO(2J+1)$ models}
\label{sub:SO}

The $SU(2)$-symmetric spin-$1$ AKLT chain has been generalized in a number of different 
directions: higher $SU(2)$-spins \cite{arovas:1988}, higher lattice 
dimensions \cite{kennedy:1988}, $SU(N)$-invariant models with $N\geq 2$ \cite{affleck:1991}, 
and most recently to a class of spin-$J$ models with an $SO(2J+1)$-invariant nearest-neighbor
interaction \cite{SOn}. The simplest way to introduce the latter generalization is to note
that the kernel of the  interaction of the AKLT model, i.e., the ground state space of two neighboring
spins is spanned by the antisymmetric vectors, namely the spin $1$ triplet, and spin singlet state, 
which is symmetric. As shown in \cite{FNW_ff}, The antisymmetric subspace of
two spins enhanced with one, arbitrary, symmetric state, is the ground state space of a 
frustration-free spin chain with a  translation invariant nearest-neighbor interaction.
In particular, the spin-$J$ chain, with integral $J\geq1$, with the nearest-neighbor interaction
given by the projection onto the orthogonal complement of the span of the antisymmetric vectors 
and the (symmetric) spin singlet state is a frustration-free model with a unique gapped MPS ground
state of the infinite chain. It is easy to see that this interaction commutes with
$SO(2J+1)$ acting by its fundamental representation on each spin.

In the case of the AKLT model ($J=1$), the antisymmetry of the ground states is reflected in the fact that the 
matrices~(\ref{v0s}, \ref{vpms}) at $s=s_0$ can be related to the generators of the Clifford 
algebra~$\caC_3$:
\begin{eqnarray}
Z_1 &=& \sqrt{3/2}(v_+-v_-),\: Z_2 = -\sqrt{3/2}\,\iu(v_++v_-)\label{CliffordCAR} \\
Z_0 &=& -\sqrt{3}v_0\,,\nonumber
\end{eqnarray}
In fact, the transformation $\{v_-,v_0,v_+\}\rightarrow\{Z_0,Z_1,Z_2\}$ corresponds, up to a globl rescaling by $\sqrt{3}$, to a change of basis in the local Hilbert space of the spins. Similarly for $J>1$, the ground states of the $SO(2J+1)$ model are, up to an overall normalization factor,
generated by the matrices $\{Z_\alpha = Z_\alpha^*:0\leq \alpha\leq2J\}$ of the higher dimensional Clifford algebra $\caC_{2J+1}$ satisfying the anticommutation relations
\begin{equation}\label{Clifford}
Z_\alpha Z_\beta + Z_\beta Z_\alpha = 2\delta_{\alpha\beta} 1.
\end{equation}
The symmetry of the ground state is manifest in the invariance of~(\ref{Clifford}) under the transformation $Z'_\beta = \sum_{\alpha} O_{\beta\alpha} Z_\alpha$ for any orthogonal matrix $O$. The normalization factor $\gamma$ can be found by setting $\sum_\alpha \gamma^2Z_\alpha^2=1$,
yielding $\gamma = (2J+1)^{-1/2}$.

A representation of the Clifford 
algebra $\caC_{2J+1}$ can be obtained from a representation of the algebra of canonical anticommutation relations (CAR) with $J$ creation operators
\begin{equation*}
a\str_j = \underbrace{(-1)^q\otimes \cdots \otimes (-1)^q}_{j-1} \otimes \sigma^+ \otimes \underbrace{ 1  \otimes \cdots \otimes  1 }_{J-j}\,.
\end{equation*}
where $q = 1-p = \sigma^-\sigma^+$. Note that $(-1)^q = \sigma^3$. Then, for $1\leq j\leq J$,
\begin{equation*}
Z_{2j-1} = a_j + a\str_j\,,\qquad Z_{2j} = \iu(a_j - a\str_j)\,,
\end{equation*}
and
\begin{equation*}
Z_0 = \prod_{j=1}^J (2a\str_j a_j - 1) = \bigotimes_{j=1}^J(-1)^q\,.
\end{equation*}
The canonical anticommutation relations of the $(a\str_j, a_j)_{j=1}^J$ imply the Clifford relations for the $(Z_\alpha)_{\alpha=0}^{2J}$. This representation is of dimension $2^J$.

Now, the matrices
\begin{equation*}
{V}_{2j-1} = \alpha_ja_j\qquad {V}_{2j} = \beta_ja_j\str\qquad {V}_0 = \gamma Z_0
\end{equation*}
are related to the normalized Clifford generators by a change of basis if and only if $\abs{\alpha_j}^2 = \abs{\beta_j}^2 = 2/(2J+1)$. In that case, the set $\{V_0\}\cup\{V_{2j-1},V_{2j}:1\leq j \leq J\}$ generate the same matrix product states as the $(Z_\alpha)_{\alpha=0}^{2J}$, namely the ground states of the $SO(2J+1)$ invariant model. Note that $\sum_j \abs{\alpha_j}^2 + \abs{\gamma}^2 = 1$, so that it is natural to set
\begin{equation*}
\gamma =: \cos(s_0),\quad\text{and}\quad\alpha_j = \frac{1}{\sqrt{J}} \sin(s_0).
\end{equation*}
for all $j=1,\ldots, J$.

It remains to introduce a deformation of the CAR algebra to relate it to the PVBS algebra with $n_L = n_R = J$. We first note that for any complex number $\lambda$,
\begin{equation}
\lambda \cdot \sigma^- \lambda^q =  \lambda^q \sigma^- \,,\qquad \sigma^+ \lambda^q = \lambda \cdot \lambda^q \sigma^+,
\label{twistedCAR_2}
\end{equation}
where $\lambda^{q}=p + \lambda q$.
For parameters $\lambda_1,\ldots,\lambda_J$, we introduce the following twisted creation operators
\begin{equation*}
a\str_j(\lambda) = (-\lambda_1)^q\otimes \cdots \otimes (-\lambda_{j-1})^q \otimes \sigma^+ \otimes \lambda_{j+1}^q \otimes \cdots \otimes \lambda_{J}^q\,,
\end{equation*}
their adjoints, and
\begin{equation*}
a_0(\lambda) = (-\lambda_1)^q \otimes \cdots \otimes (-\lambda_J)^q \,.
\end{equation*}
Clearly, $a\str_j(\lambda)^2 = 0 = a_j(\lambda)^2$. A direct consequence of~(\ref{twistedCAR_2}) are
the twisted commutation relations:
\begin{eqnarray*}
a_j\str(\lambda) a_j(\lambda) + \lambda_j^2 a_j(\lambda) a_j\str(\lambda) &=& a_0(\lambda)^2 \label{TwComm1}\\
a\str_j(\lambda) a_k(\lambda) + \lambda_j \lambda_k a_k(\lambda) a\str_j (\lambda)&=& 0 \qquad (j\neq k) \label{TwComm2}\\
a_j\str(\lambda) a_0(\lambda) + \lambda_j a_0(\lambda) a_j\str(\lambda) &=&0 \label{TwComm3}\\
a\str_j(\lambda) a\str_k(\lambda) + \lambda_j \lambda_k^{-1} a\str_k(\lambda) a\str_j(\lambda) &=&0 \label{TwComm4}
\end{eqnarray*}
and their adjoint relations. The higher dimensional analog of the algebraic path~(\ref{v0s}, \ref{vpms}) is easily obtained with the following definitions for $s\in[0,s_0]$:
\begin{equation*}
{V}_{2j-1}(s) = \alpha_j(s) a_j(s), \quad {V}_{2j}(s) = \beta_j(s)a_j\str(s)
\end{equation*}
and ${V}_0(s) = \gamma(s) a_0(s)$, where $a_j^\sharp(s) = a_j^\sharp(\lambda(s))$ for a smooth path of parameter vector $\lambda(s)$ such that $\lambda_j(s)<1$ for $s<s_0$ and $\lambda_j(s_0)=1$. Using the commutation relations, the normalization $\sum_{\alpha}V_\alpha\str V_\alpha=1$ reads
\begin{equation*}
 1  = \sum_{j=1}^J\left(\abs{\beta_j}^2 - \lambda_j^2 \abs{\alpha_j}^2 \right)a_j a_j\str + \Big[\sum_{j=1}^J \abs{\alpha_j}^2 + \abs{\gamma}^2\Big]a_0^2
\end{equation*}
which implies 
\begin{equation*}
\sum_{j=1}^J \abs{\alpha_j}^2 + \abs{\gamma}^2 = 1 \quad\text{and}\quad \abs{\beta_j}^2 = 1- \lambda_j^2 (1-\abs{\alpha_j}^2).
\end{equation*}
Note that the $SO(2J+1)$ symmetry is broken as soon as $\abs{\beta_j}\neq\abs{\alpha_j}$. 
Concretely we choose $\gamma(s) = \cos(s)$ and $\alpha_j(s) = \sin(s)/\sqrt{J}$ for $s\in[0,s_0]$, 
thereby producing a path that mimics~(\ref{v0s}, \ref{vpms}). The commutation relations of the 
$V_\alpha$'s interpolate between those of the CAR (and therefore of the $SO(2J+1)$ models) and of the PVBS with parameters
\begin{equation} \label{PVBS_Parameters}
\{\lambda_j(0),\lambda_j(0)^{-1}:1\leq j\leq J\};
\end{equation}
In particular,
\begin{multline*}
V_{2j}(s)V_{2j-1}(s) + \lambda_j(s)^2 V_{2j-1}(s)V_{2j}(s) \\ = \frac{\sin(s)\left[1- \lambda_j^2 (1-\sin^2(s)/J)\right]}{\sqrt{J}\cos^2(s)}V_0^2\,.
\end{multline*}

This algebraic path of matrices generates a smooth path of matrix product states, from which a smooth family of Hamiltonians with nearest-neighbor interaction can be constructed, for any $J$. They interpolate between the $SO(2J+1)$ model and the PVBS model with $n_L = n_R = J$. The properties needed to conclude that the gap remains open along the path can be derived from a suitable generalization of the arguments for the AKLT path, $J=1$, and will be included in future work.


\section{Discussion}
\label{sec:discussion}

The fact that one of each of the $n$ types of particles can appear in a ground state of
the PVBS system, i.e., without raising the energy, should not be interpreted as implying that the
particles are massless. Quite to the contrary, there is a mass gap for each of them. It turns out,
however, that they can bind to the left or right edge of the chain and the binding energy exactly 
equals the mass gap so that such states with a particle bound to the edge are degenerate with the 
vacuum ground state.

As we explained in the previous section, the AKLT model is equivalent to a PVBS model
with two types of particles, one that binds to the left edge and one that binds to the right
edge. For a finite chain, this yields a four-dimensional ground state space (no particles; one
particle on the left boundary; one particle on the right boundary; two particles, one on the left
and one on the right). The bulk ground state is unique. It has a finite correlation
length in the AKLT model but is a simple product state in the equivalent PVBS model.
The unitary transformation relating the two models is quasi-local~\cite{BMNS}, 
meaning that, under conjugation
with the unitary, local observables map to observables that are approximately local, i.e., that
up to an arbitrarily small correction depend only on a finite number of spins. Since our goal is to understand the structure of bulk phases and edge states (and, in higher
dimensions also the effects of topology), the quasi-locality of this unitary transformation is 
essential.
In finite volume, any unitary will preserve the dimension of the ground state space. The importance of the locality property comes into sharp focus when one takes 
the thermodynamic limit and considers the resulting ground states on the infinite and half-infinite 
chains. For 
example, the non-local unitary transformation introduced by Kennedy and Tasaki 
\cite{Hidden1,*Hidden} to 
reveal the hidden string order \cite{String} transforms the 4 AKLT  ground states into 4 
translation invariant product states, which leads to 4 distinct bulk ground states in the 
thermodynamic limit. This example shows that non-local unitary transformations do not preserve
the structure of the bulk ground state(s) and clearly would not be useful to classify gapped
ground state phases.

The bulk ground state of a $SO(2J+1)$ model for any integer $J$ is equivalent to a unique 
product state. This however is not sufficient for them to all belong to the same gapped phase as the 
number of edge states depends on $J$. The definition of a phase through a quasi-local 
automorphism yields a finer classification that takes these boundary states into account.

Several generalizations of the PVBS are possible. First, it is straightforward to describe 
models with more than one vacuum state, e.g., models with a broken discrete symmetry. 
This could include breaking of the translation symmetry of the lattice to a subgroup, leading to
periodic ground states. Second, the construction can easily be generalized to allow more than 
one particle to occupy the same site. The auxiliary chain of $n$ spin-$1/2$ particles we have
used in the appendix just has to be replaced by an arbitrary finite spin chain.
Third, if all or a subset of the $\lambda$-parameters are equal, one can describe
Hamiltonians with a local continuous symmetry group. Adding symmetry as a constraint
considerably enriches the classification problem of the gapped ground state phases and 
the phase transitions between them \cite{Tu_etal,WenSymm}. 

It appears to us that the class of models with PVBS ground states is able to capture the
range of behaviors seen in gapped ground states of one-dimensional spin systems
with short range interactions. The most interesting phenomena from the physical point of view
occur in two dimensions. Some aspects of the classification problem have already been 
considered in two and higher dimensions. The notion of automorphic equivalence studied in
\cite{BMNS}, e.g., applies to arbitrary dimensions. Interesting boundary states in two dimensions 
have been studied in \cite{WenEdgeHall, *Zuri}. Much further work is needed to better understand the landscape of gapped ground state phases in two and three dimensions. 


\bibliography{Ref_pvbs}

\begin{acknowledgments} The authors gratefully acknowledge the kind hospitality of the Erwin Schr\"odinger International Institute for Mathematical Physics (ESI) in Vienna, Austria, where part of the work reported here was carried out. This work was also supported by the National Science Foundation: S.B. under Grant \#DMS-0757581 and B.N. under grant
\#DMS-1009502
\end{acknowledgments}
%
%
%
%
%
%
\end{document}